\documentclass[12pt]{article}

\usepackage{amssymb}
\usepackage{graphicx}
\usepackage{subfigure}
\usepackage{hyperref}
\usepackage{mathrsfs}
\usepackage{latexsym}
\usepackage{amsfonts}
\usepackage{amsmath}
\usepackage{esint}
\usepackage{amsxtra}
\usepackage[enableskew]{youngtab}

\newdimen\mathindent
\mathindent\leftmargini

\def \ep1{\epsilon_1}
\def \ep2{\epsilon_2}
\def \m{\mbox}

\def \be{\begin{equation}}
\def \ee{\end{equation}}
\def\beq{\begin{eqnarray}}
\def\eeq{\end{eqnarray}}
\def \ba{\begin{array}}
\def \ea{\end{array}}
\def\nn{\nonumber}
\def \exp{\mbox{exp}}
\def \f{\frac}

\def \p{\partial}
\def \<{\langle}
\def \>{\rangle}
\def \wt{\widetilde}
\def \wh{\widehat}

\def \Y{\widetilde{Y}}
\def \rp{|v_0\rangle}
\def \lp{\langle v_0|}

\begin{document}
\vspace*{-.6in}
\thispagestyle{empty}
\baselineskip = 18pt

\vspace{.5in}
\vspace{.5in}
{\LARGE
\begin{center}
A note on W symmetry of $N=2$ gauge theory
\end{center}}

\vspace{1.0cm}

\begin{center}

Wei He\footnote{weihe@itp.ac.cn}\\
\vspace{1.0cm}\emph{Center of Mathematical Sciences, Zhejiang
University, Hangzhou 310027, China}
\end{center}
\vspace{1.0cm}

\begin{center}
\textbf{Abstract}
\end{center}
\begin{quotation}
\noindent  The AGT correspondence indicates $\mathcal{N}=2$ gauge
theory possesses of W algebra symmetry. We study how the conformal
block of Toda CFT gives the expectation value of Casimir operators
of gauge theory. The $A_2$ Toda CFT with $W_3$ symmetry is taken as
the main example.
\end{quotation}

\pagenumbering{arabic}

\newpage

\section{Introduction}

Some recent study reveals interesting relations between different
subjects: the 4D gauge theory, 2D CFT, integrable theory, quantum
algebra, etc. The Alday-Gaiotto-Tachikawa(AGT)\cite{agt}
correspondence, and its generalization by Wyllard\cite{wyllard},
establishes a relation between $\mathcal{N}=2$ supersymmetric gauge
theory in the $\Omega$ background and Liouville/Toda conformal field
theory. The Nekrasov-Shatashvili(NS)\cite{NS0908} correspondence
relates the same $\mathcal {N}=2$ gauge theory with certain quantum
integrable models. All these subjects have some structures that were
mostly studied in their own field before the recent discoveries, now
it is an interesting problem to study how these properties are
realized in each related subjects, and hope to gain some new
insights.

At the moment, in the context of AGT we have explicit examples where
the gauge theory partition function is identified with the CFT
conformal block\cite{agt,wyllard,g0908}, the nonlocal operators of
gauge theory are realized by CFT operators\cite{loop1, loop2, loop3,
loop4}, and some other developments. In the context of NS we also
have examples where the Bethe equation and Hamiltonians of
integrable models are derived from the gauge theory partition
function\cite{NS0908}. In this paper, through some simple examples,
we show how the Hamiltonians can be obtained from CFT conformal
block. Although there are evidences that the whole class of
$\mathcal {N}=2$ theories are quantum integrable, we focus on
particular theories, and it is enough to demonstrate the basic
properties. The first is the SU(N) pure gauge theory, with the dual
irregular conformal block of $A_{N-1}$ Toda CFT\cite{g0908}, and in
the semiclassical limit(or the minisuperspace limit of CFT) related
to the quantized periodic Toda chain\cite{NS0908}. The second one is
the $\mathcal {N}=2^*$ gauge theory, the related CFT is the Toda CFT
on torus, and they are related to the periodic elliptic
Calogero-Moser model. For technical reasons, SU(3) gauge theory and
the related $W_3$ algebra are our main examples.

In the next section we briefly review the Gaiotto's irregular
conformal block of Liouville CFT\cite{g0908}, its relation to the
Gram/Shapovalov matrix. In section 3, 4 and 5 we show how the SU(3)
gauge theory calculation and $W_3$ algebra of $A_2$ Toda CFT can be
consistent with each other and hence with the requirements of
quantum integrable chains.

\section{Irregular conformal block and Gaiotto state}

For SU(2) pure gauge theory, its Seiberg-Witten curve\cite{sw1, sw2}
can be written in the form\cite{g0908} \be x^2-\phi_2(z)=0,\qquad
\phi_2(z)=\f{\Lambda^2}{z^3}+\f{2u}{z^2}+\f{\Lambda^2}{z}.\label{su2curve}\ee
where $u\sim<\m{tr}\varphi^2>$ is the Coulomb parameter, $\varphi$
is the adjoint scalar in the vector supermultiplet. In the AGT
correspondence, the Seiberg-Witten curve is quantized and promoted
to the operator equation $x^2-\wh{\phi}_2(z)=0$\cite{agt}. The
operators are understood as correlator of the corresponding
Liouville CFT. The operator $\wh{\phi}_2(z)$ is identified with the
energy-momentum operator of the CFT:
$\wh{\phi}_2(z)=T(z)=\sum_nL_nz^{-n-2}$. Then (\ref{su2curve})
indicates there exists a state $|G\>$ that $L_1|G\>=\Lambda^2|G\>,
L_n|G\>=0, n\ge2$, where $|G\>$ is the Gaiotto state of the
Liouville CFT, a kind of coherent state, first constructed by
Gaiotto in\cite{g0908}. It can be expanded according to the level:
$|G\>=|v_0\>+\sum_{k=1}^\infty\Lambda^{2k}|v_k\>$, where the CFT
expansion parameter $\Lambda$ is identified with gauge theory scale,
and $|v_0\>$ is the highest weight state: $L_0|v_0\>=\Delta|v_0\>,
L_n|v_0\>=0, n\ge1$. At each level $k$ the vector $v_k$ is linear
combination of descendant states of the form $L_{-k_l}\cdots
L_{-k_2}L_{-k_1}|v_0\>$ with $k_l\ge k_{l-1}\ge\cdots\ge k_1>0$ and
$\sum k_i=k$. Gaiotto has given the explicit forms of $|G\>$ at the
first few level, as later pointed out in\cite{mmm}, it can be
systematically constructed from the Gram/Shapovalov matrix of the
Liouville CFT.

If we associate the partition $(k_l,\cdots,k_2,k_1)$ with a Young
diagram $Y$, then the vector $L_{-k_l}\cdots L_{-k_2}L_{-k_1}|v_0\>$
at level $k$ can be denoted as $L_{-Y}|v_0\>$. Then the level $k$
Gram/Shapovalov matrix is $K_{[k]}(\Y;Y)=\<v_0|L_{\Y}L_{-Y}|v_0\>$,
and the coefficients of the linear combination at level $k$ turn out
to be elements of the inverse Gram/Shapovalov matrix at level
$k$\cite{mmm}:\be
|v_k\>=\sum_{|Y|=k}K_{[k]}^{-1}([1^k];Y)L_{-Y}|v_0\>.\ee where
$\Y=[1^k]$ corresponds to vector $L_{-1}^k|v_0\>$. The norm of
$|v_k\>$ is \be \<
v_k|v_k\>=\sum_{|Y|=|\Y|=k}K_{[k]}^{-1}([1^k];\Y)K_{[k]}^{-1}([1^k];Y)\<v_0|
L_{\Y}L_{-Y}|v_0\>=K_{[k]}^{-1}([1^k];[1^k]).\ee In order to relate
CFT data to gauge theory data, we identify the CFT parameters with
the gauge theory parameters as $c=1+6Q^2(\epsilon_1\epsilon_2)^{-1},
\Delta=(Q^2/4-a^2)(\epsilon_1\epsilon_2)^{-1}$, where
$Q=\epsilon_1+\epsilon_2$ is the background charge of CFT, then
$\<v_k|v_k\>$ equals to the Nekrasov's $k$-th instanton partition
function $Z_k$ of $\mathcal {N}=2$ gauge theory\cite{instcount}. The
norm of the Gaiotto state therefore equals to the full instanton
partition function $Z^{inst}(a,q)$.

In fact, (\ref{su2curve}) also indicates the following relation,\be
\f{\<G|L_0|G\>}{\<G|G\>}=2u.\ee This is very easy the verify. Using
the Virasoro relation(see (\ref{w3alg}) in section four), we have
$L_0L_{-k_l}\cdots L_{-k_2}L_{-k_1}|\Delta\>=(\Delta+k)|\Delta\>$.
Therefore\beq \f{\<G|L_0|G\>}{\<G|G\>}&=&\f{\sum_{k=0}^\infty\<
v_k|\Delta+k|v_k\>\Lambda^{4k}}{\sum_{k=0}^\infty\<v_k|v_k\>
\Lambda^{4k}}=\Delta+\Lambda^4\f{\p}{\p\Lambda^4}\m{ln}Z^{inst}=2u\eeq
If we make the relation between $u$ and $<\m{tr}\varphi^2>$ precise,
it is
$2u=(-\epsilon_1\epsilon_2)^{-1}(<\m{tr}\varphi^2>-\f{1}{4}Q^2)$,
then the relation above is the Matone's relation of $\mathcal {N}=2$
gauge theory\cite{matone, ffmp}.

The above is the story for the SU(2) pure gauge theory and its
irregular Liouville conformal block. Generalization of this
construction to other cases has been studied, including theories
with mass deformations, with other gauge groups, see for
example\cite{g0908, mm, t0912, kms, kmst}. For example, the pure
SU(N) gauge theory should be related to irregular conformal block of
the $A_{N-1}$ Toda CFT, the corresponding conformal algebra is the
$W_N$ algebra\cite{bs9210}. The Seiberg-Witten curve can be written
as\cite{g0904, gmn}\be
x^N-\phi_2(z)x^{N-2}-\phi_3(z)x^{N-3}-\cdots-\phi_N(z)=0,\nn\ee
with\be \phi_s(z)=\f{2u_s}{z^s},\qquad s=2,3,\cdots N-1, \qquad
\phi_N(z)=\f{\Lambda^N}{z^{N+1}}+\f{2u_N}{z^N}+\f{\Lambda^N}{z^{N-1}}.\label{SWcurve}\ee
here $u_s\sim\m{tr}\varphi^s$ are the Coulomb parameters. After
promoting this relation to operator relation of Toda CFT, the
operators $\wh{\phi}_s(z)$ are identified with spin-$s$ currents
$W^{(s)}(z)$ of the CFT\cite{bt0909, kmst0911}, among of them
$W^{(2)}(z)=T(z)$ is the energy-momentum current. As the mode
expansion of currents are
$W^{(s)}(z)=\sum_{n\in\mathbb{Z}}W^{(s)}_nz^{-(n+s)}$, then the
Gaiotto states can be consistently constructed from the constraints,
\beq W^{(s)}_n|G\>&=&0,\qquad
n\ge1,\qquad s=2,3\cdots, N-1,\nn\\
W^{(N)}_1|G\>&=&\Lambda^N|G\>,\qquad W^{(N)}_n|G\>=0,\qquad
n\ge2.\label{LWconstr}\eeq It is shown in \cite{kmst} that at each
level the coefficients of linear combination are elements of the
inverse Gram/Shapovalov matrix, and $\<v_k|v_k\>$ equals to the
element $K_{[k]}^{-1}((W_{-1}^{(N)})^k,W_{-1}^{(N)})^k)$, and should
equal to the instanton partition function $Z_k$, if the CFT
parameters are properly identified with gauge theory parameters,\be
\<G|G\>=Z^{inst},\qquad \m{i.e.}\quad \<v_k|v_k\>=Z_k,\quad
k\ge1,\label{parti}\ee We should also expect the following relation
holds, \be \f{\<G|W^{(s)}_0|G\>}{\<G|G\>}=2u_s,\qquad s=2,3\cdots
N.\label{hamil}\ee For various case, it has been shown in\cite{mm,
t0912, kms, kmst, bmt, kt1203} by some explicit calculations that
the Gaiotto state of Toda CFT indeed gives gauge theory instanton
partition. In the case of pure gauge theory, it is simply the norm
of $|G\>$ as in(\ref{parti}), and in fact it is indeed equal to a
particular element of the inverse Gram/Shapovalov matrix. In the
next two section, we will show the $W_3$ CFT also gives Casmirs of
SU(3) pure gauge theory as in(\ref{hamil}), and in fact also
directly related to elements of the inverse Gram/Shapovalov matrix.
The $N=2^*$ gauge theory and the CFT on the torus are also
discussed.

\section{Instanton calculation for SU(3) gauge theory}

The Nekrasov instanton partition function for $k$-instanton sector
of SU(N) pure gauge theory can be evaluated by the
formula\cite{fp0208}:
 \be Z_k=\sum_{\lbrace
\sum|Y_\alpha|=k\rbrace}\prod_{\alpha,\beta=1}^N\prod_{s\in
Y_\alpha}\prod_{s^{'}\in Y_\beta}
\f{1}{E_{\alpha\beta}(s)(\epsilon_+-E_{\beta\alpha}(s^{'}))},\ee
with $\epsilon_+=\epsilon_1+\epsilon_2$, and  \be
E_{\alpha\beta}(s)=a_{\alpha\beta}-h_\beta(s)\epsilon_1+(v_\alpha(s)+1)\epsilon_2,\ee
the v.e.v of the adjoint scalar satisfy $\sum_{i=1}^Na_i=0$. The sum
is running over all possible partitions symbolled by the Young
diagrams $\lbrace Y_\alpha\rbrace$ with $\sum|Y_\alpha|=k$. (These
Young diagrams have no direct relation with Young diagrams
associated to descendant states on the CFT side.) Moreover, the
expectation value of the Casimirs $\m{tr}\varphi^m$ can be
calculated by\cite{mn, ny0311}\be
<\m{tr}\varphi^m>=\f{1}{Z^{inst}}\sum_{k=0}^\infty\sum_{\lbrace
\sum|Y_\alpha|=k\rbrace}\f{\m{ch}_m(\lbrace
Y_\alpha\rbrace)}{\prod_{\alpha,\beta=1}^N\prod_{s\in
Y_\alpha}\prod_{s^{'}\in
Y_\beta}E_{\alpha\beta}(s)(\epsilon_+-E_{\beta\alpha}(s^{'}))}\Lambda^{2kN}.\label{phivev}\ee
where $q_{in}=\Lambda^{2N}$ is the instanton expansion parameter,
and \beq \m{ch}_m(\lbrace
Y_\alpha\rbrace)&=&\f{1}{m!}\sum_{\alpha=1}^N\lbrace
a_\alpha^m-\sum_{s(i_\alpha,j_\alpha)\in Y_\alpha}\lbrack
(a_\alpha+j_\alpha\epsilon_1+i_\alpha\epsilon_2)^m-(a_\alpha+j_\alpha\epsilon_1+(i_\alpha-1)\epsilon_2)^m\nn\\
&\quad&-(a_\alpha+(j_\alpha-1)\epsilon_1+i_\alpha\epsilon_2)^m
+(a_\alpha+(j_\alpha-1)\epsilon_1+(i_\alpha-1)\epsilon_2)^m\rbrack\rbrace.\eeq
It can be read from the degree $m$ part of the Chern character of
the equivalent bundle $\m{Ch}_{\vec{Y}}(\mathcal{E})$, \be
\m{ch}_m(\lbrace
Y_\alpha\rbrace)=\sum_{\alpha}^N[e^{a_\alpha}-(1-e^{\epsilon_1})(1-e^{\epsilon_2})\sum_{s(i_\alpha,j_\alpha)\in
Y_\alpha}e^{a_\alpha+(j_\alpha-1)\epsilon_1+(i_\alpha-1)\epsilon_2}]|_{m}.\ee
In the NS correspondence, $<\m{tr}\varphi^m>$ as functions of the
quasimomenta $a_i$, are the quantized Hamiltonians of Toda chain.
The quasimomenta $a_i$ are constrained by the Bethe equation that
determine the critical points of the gauge theory
prepotential\cite{NS0908}.

For SU(3) theory, the nontrivial independent Casimires are
$\m{tr}\varphi^2$ and $\m{tr}\varphi^3$. It is simple to see for
$m=2$ we have $\m{ch}_2(\lbrace Y_\alpha\rbrace)=\f{1}{2}\sum_\alpha
a_\alpha^2-k\epsilon_1\epsilon_2$, then (\ref{phivev}) leads to the
Matone's relation, the information of $<\m{tr}\varphi^2>$ can be
derived from the partition function. This conclusion is valid for
SU(N) theory\cite{ffmp}. More information can be found in
$<\m{tr}\varphi^3>$, \be\m{ch}_3(\lbrace
Y_\alpha\rbrace)=\f{1}{6}\sum_{\alpha=1}^N\lbrace
a_\alpha^3-3\epsilon_1\epsilon_2\sum_{s\in
Y_\alpha}[2a_\alpha+(2j_\alpha-1)\epsilon_1+(2i_\alpha-1)\epsilon_2]\rbrace.\ee

Let us denote\be \m{tr}\varphi^m_{k}=\sum_{\lbrace
\sum|Y_\alpha|=k\rbrace}\f{\m{ch}_m(\lbrace
Y_\alpha\rbrace)}{\prod_{\alpha,\beta=1}^N\prod_{s\in
Y_\alpha}\prod_{s^{'}\in
Y_\beta}E_{\alpha\beta}(s)(\epsilon_+-E_{\beta\alpha}(s^{'}))},\ee
Then with no instanton correction, $k=0$, we have \be
\m{tr}\varphi^3_{k=0}=\f{1}{2}(a_1^3+a_2^3+a_3^3)=-\f{1}{2}a_1a_2(a_1+a_2).\ee
For one instanton correction, $k=1$, we have contributions from the
partitions $([1],\emptyset,\emptyset)$, and from its permutations
obtained by $a_1\leftrightarrow a_2, a_1\leftrightarrow a_3$. They
contribute \beq
\m{tr}\varphi^3_{k=1}&=&\f{1}{6}\sum_{\alpha=1}^3\f{a_1^3+a_2^3+a_3^3-3\epsilon_1\epsilon_2(2a_\alpha+\epsilon_+)}{\epsilon_1\epsilon_2\prod_{\beta\ne\alpha}^3a_{\alpha\beta}(a_{\alpha\beta}+\epsilon_+)}
\nn\\&=&-\f{1}{2}a_1a_2(a_1+a_2)Z_1-\f{9a_1a_2(a_1+a_2)}{((a_1-a_2)^2-\epsilon_+^2)((2a_1+a_2)^2-\epsilon_+^2)((a_1+2a_2)^2-\epsilon_+^2)}.\eeq
For two instanton correction, $k=2$, we have partitions
$([1],[1],\emptyset),([11],\emptyset,\emptyset),([2],\emptyset,\emptyset)$,
and their permutations. They contribute \be
\m{tr}\varphi^3_{k=2}=-\f{1}{2}a_1a_2(a_1+a_2)Z_2-\f{27a_1a_2(a_1+a_2)(a_1^2+a_1a_2+a_2^2-\epsilon_+^2+\epsilon_1\epsilon_2)
(8a_1^6+24a_1^5a_2+\cdots)}{\epsilon_1\epsilon_2\prod_{\alpha<\beta}(a_{\alpha\beta}^2-\epsilon_+^2)(a_{\alpha\beta}^2-(\epsilon_++\epsilon_1)^2)(a_{\alpha\beta}^2-(\epsilon_++\epsilon_2)^2)}\ee

\subsection{$\mathcal{N}=2^*$ gauge theory}

The $\mathcal{N}=2^*$ gauge theory couples an adjoint matter to the
vector multiplet. If the physical mass is $m^*$, then it is the
parameter $m=m^*+(\epsilon_1+\epsilon_2)/2$ appears in the Nekrasov
instanton partition function. It can be evaluated by the
formula\cite{bfmt}, \be
Z_k=\sum_{\sum|Y_\alpha|=k}\prod_{\alpha,\beta=1}^{N}\prod_{s\in
Y_\alpha}\prod_{s^{'}\in
Y_\beta}\f{(E_{\alpha\beta}(s)-m)(\epsilon_+-E_{\beta\alpha}(s^{'})-m)}
{E_{\alpha\beta}(s)(\epsilon_+-E_{\beta\alpha}(s^{'}))},\label{adjointcount}\ee
For SU(3) theory, the one instanton contribution gives the partition
function\be
Z_1=\f{3(m-\epsilon_1)(m-\epsilon_2)(4a_1^6+12a_1^5a_2-3a_1^4a_2^2+\cdots+2m\epsilon_2^5-6\epsilon_1\epsilon_2^5-\epsilon_2^6)}
{\epsilon_1\epsilon_2((a_1-a_2)^2-\epsilon_+^2)((2a_1+a_2)^2-\epsilon_+^2)((a_1+2a_2)^2-\epsilon_+^2)}.\ee

The the expectation value of Casimirs of $\mathcal{N}=2^*$ theory is
give by a formula similar to(\ref{phivev}), now with\be
\m{tr}\varphi^m_{k}=\sum_{\lbrace
\sum|Y_\alpha|=k\rbrace}\left(\prod_{\alpha,\beta=1}^N\prod_{s\in
Y_\alpha}\prod_{s^{'}\in
Y_\beta}\f{(E_{\alpha\beta}(s)-m)(\epsilon_+-E_{\beta\alpha}(s^{'})-m)}{E_{\alpha\beta}(s)(\epsilon_+-E_{\beta\alpha}(s^{'}))}\right)\m{ch}_m(\lbrace
Y_\alpha\rbrace).\label{phivevstar}\ee $<\m{tr}\varphi^2>$ leads to
the Matone's relation as pure gauge theory case. The one instanton
result for $<\m{tr}\varphi^3>$ is \be
\m{tr}\varphi^3_{k=1}=-\f{1}{2}a_1a_2(a_1+a_2)Z_1-\f{3(m-\epsilon_1)(m-\epsilon_2)
(6m^4a_1^2a_2+6m^4a_1a_2^2+\cdots-7\epsilon_1\epsilon_2^6-\epsilon_2^7)}{2((a_1-a_2)^2-\epsilon_+^2)((2a_1+a_2)^2-\epsilon_+^2)((a_1+2a_2)^2-\epsilon_+^2)}.\ee
Let us denote the coupling of $\mathcal{N}=2^*$ theory by $q$, then
in the decoupling limit $q\to0, m\to\infty$, while keep
$qm^6=\Lambda^6$, the results for $N=2^*$ theory reduce to that for
pure gauge theory.

\section{Irregular conformal block for A$_2$ Toda CFT}

According to Wyllard's generalization\cite{wyllard} of the the
AGT\cite{agt}, SU(3) gauge theories are related to the $A_2$ Toda
CFT on surfaces with punctures, the underlying conformal algebra is
the $W_3$ algebra, generated by the spin two current
$W^{(2)}=T(z)=\sum_{n\in\mathbb{Z}}L_nz^{-n-2}$ and the spin three
current $W^{(3)}=W(z)=\sum_{n\in\mathbb{Z}}W_nz^{-n-3}$. The $W_3$
algebra is
\beq &\quad&\lbrack L_m, L_n\rbrack=(m-n)L_{m+n}+\f{c}{12}m(m^2-1)\delta_{m+n,0},\nn\\
&\quad&\lbrack L_m, W_n\rbrack=(2m-n)W_{m+n},\nn\\
&\quad&\lbrack W_m, W_n\rbrack=\f{9}{2}\lbrace\f{c}{3\cdot
5!}m(m^2-1)(m^2-4)\delta_{m+n,0}+(m-n)[\f{16}{22+5c}\Lambda_{m+n}\nn\\
&\quad&\qquad\qquad\quad+(\f{(m+n+2)(m+n+3)}{15}-\f{(m+2)(n+2)}{6})L_{m+n}]\rbrace.\label{w3alg}\eeq
with \be \Lambda_n=\sum_{m\in
\mathbb{Z}}:L_mL_{n-m}:+\f{x_n}{5}L_n,\ee where for even $n$:
$x_{2l}=(1-l)(1+l)$, and for odd $n$: $x_{2l+1}=(1-l)(2+l)$. The
central charge is $c=2+24Q^2/(\epsilon_1\epsilon_2)$ with
$Q=\epsilon_1+\epsilon_2$ the background charge of Toda CFT.

For the pure gauge theory, there exists the $A_2$  irregular
conformal block and the related Gaiotto state. Again, the Gaiotto
state can be expanded as
$|G\>=|v_0\>+\Lambda^3|v_1\>+\Lambda^6|v_2\>+\cdots$, at each level
$|v_k\>$ is linear combination of vectors of the form
$L_{-Y_1}W_{-Y_2}\rp$ with $|Y_1|+|Y_2|=k$. Here
$L_{-Y_1}=L_{-k_l}\cdots L_{-k_2}L_{-k_1}$ and similarly
$W_{-Y_2}=W_{-k^{'}_l}\cdots W_{-k^{'}_2}W_{-k^{'}_1}$. We always
write the product of generators $L_{-n}$ on the left of the product
of $W_{-n}$. $|v_0\>$ is he highest weight state satisfying \be
L_0\rp=\Delta\rp, \qquad L_n\rp=0,\qquad n\ge1,\ee and  \be
W_0\rp=w\rp, \qquad W_n\rp=0,\qquad n\ge1,\ee

From the singular structure of the Seiberg-Witten curve of the SU(3)
pure gauge theory(\ref{SWcurve}),
the Gaiotto state should satisfy(\ref{LWconstr}) \beq L_n|G\>&=&0,\qquad n\ge1\\
W_1|G\>&=&\Lambda^3|G\>,\qquad W_n|G\>=0,\quad n\ge2.\eeq

Then we can determine $|G\>$ level by level, using its
Gram/Shapovalov matrix\cite{kmst}. The Gram/Shapovalov matrix is \be
K(\widetilde{Y}_1,\widetilde{Y}_2;Y_1,Y_2)=\lp
W_{\widetilde{Y}_2}L_{\widetilde{Y}_1}\cdot L_{-Y_1}W_{-Y_2}\rp.\ee
The first two level CFT data are given in\cite{mm, t0912}, we also
give them in the Appendix\ref{appendixI}. At the level $k=1$, we
have\be |v_1\>=\f{1}{9(D\Delta^2-w^2)}(-3w L_{-1}+2\Delta
W_{-1})\rp,\ee We have used the notations from\cite{mm},
$D=\f{32}{22+5c}(\Delta+\f{1}{5})-\f{1}{5}$. Note that the two
coefficients of $L_{-1}$ and $W_{-1}$ are
$c_1^{[1]}=K_{[1]}^{-1}(\emptyset,[1];[1],\emptyset)$ and
$c_2^{[1]}=K_{[1]}^{-1}(\emptyset,[1];\emptyset,[1])$ respectively.
Therefore $|v_1\>$ coincide with the level one null state. It can be
shown that if we make the identification\cite{mm}\be
\Delta=\f{a_1^2+a_2^2+a_1a_2-\epsilon_+^2}{-\epsilon_1\epsilon_2},\quad
w=6(\f{6}{22+5c})^{1/2}\f{a_1a_2(a_1+a_2)}{(-\epsilon_1\epsilon_2)^{3/2}},\quad
c=2-24\f{\epsilon_+^2}{-\epsilon_1\epsilon_2}.\label{paraident}\ee
then $\<v_1|v_1\>$ gives the one instanton partition function $Z_1$
of gauge theory, \be
\<v_1|v_1\>=\f{2\Delta}{9(D\Delta^2-w^2)}=K_{[1]}^{-1}(\emptyset,[1];\emptyset,[1])=(\epsilon_1\epsilon_2)^3(\f{22+5c}{6^3})Z_1=\widetilde{Z}_1.\ee
At the level $k=2$, we have\be
|v_2\>=(c_1^{[2]}L_{-2}+c_2^{[2]}L^2_{-1}+c_3^{[2]}L_{-1}W_{-1}+c_4^{[2]}W_{-2}+c_5^{[2]}W^2_{-1})\rp,\ee
where $c_i^{[2]}$ are the last column of the inverse Gram/Shapovalov
matrix
$c_i^{[2]}=c_{\tiny{Y_1,Y_2}}^{[2]}{}=K_{[2]}^{-1}(\emptyset,[1^2];Y_1,Y_2)$.
Its norm satisfies\be
\<v_2|v_2\>=K_{[2]}^{-1}(\emptyset,[1^2];\emptyset,[1^2])=(\epsilon_1\epsilon_2)^6(\f{22+5c}{6^3})^2Z_2=\widetilde{Z}_2,\ee
The level $k=3$ Gram/Shapovalov matrix is given in \cite{kms}, if we
write\beq
|v_3\>&=&(c_1^{[3]}L_{-3}+c_2^{[3]}L_{-2}L_{-1}+c_3^{[3]}L_{-1}^3+c_4^{[3]}L_{-2}W_{-1}+c_5^{[3]}L_{-1}^2W_{-1}+c_6^{[3]}L_{-1}W_{-2}\nn\\
&\quad&+c_7^{[3]}L_{-2}W_{-1}^2+c_8^{[3]}W_{-3}+c_9^{[3]}W_{-2}W_{-1}+c_{10}^{[3]}W_{-1}^3)\rp,\eeq
with the coefficients
$c_i^{[3]}=c_{\tiny{Y_1,Y_2}}^{[3]}=K_{[3]}^{-1}(\emptyset,[1^3];Y_1,Y_2)$,
then \be
\<v_3|v_3\>=K_{[3]}^{-1}(\emptyset,[1^3];\emptyset,[1^3])=(\epsilon_1\epsilon_2)^9(\f{22+5c}{6^3})^3Z_3=\widetilde{Z}_3,\ee

Now let us check (\ref{hamil}). First£¬ look at $\<G|L_0|G\>$. From
the $W_3$ algebra (\ref{w3alg}), $[L_0,L_{-n}]=nL_{-n},
[L_0,W_{-n}]=nW_{-n}$ it is easy to see
$L_0L_{-Y_1}W_{-Y_2}\rp=(\Delta+k)L_{-Y_1}W_{-Y_2}\rp$ if
$|Y_1|+|Y_2|=k$. Therefore, $L_0|v_k\>=(\Delta+k)|v_k\>$. The
precise relation between $u$ and $<\m{tr}\varphi^2>$ for SU(3)
theory is $2u=(-\epsilon_1\epsilon_2)^{-1}(<\m{tr}\varphi^2>-Q^2)$,
then the Matone's relation follows as in the Liouville case.

Then, let us look at $\<G|W_0|G\>$. Now apply the operator $W_0$ on
$|v_k\>$ always results in $W_0|v_k\>=w|v_k\>+|\wh{v}_k\>$ with
$|\wh{v}_k\>$ a vector also at the level $k$. For example,\beq
|\wh{v}_1\>&=&(\f{9}{2}Dc_2^{[1]}L_{-1}+2c_1^{[1]}W_{-1})|v_0\>,\nn\\
|\wh{v}_2\>&=&[\f{288\Delta}{22+5c}c_4^{[2]}L_{-2}+(\f{9}{2}Dc_3^{[2]}+\f{144}{22+5c}c_4^{[2]})L_{-1}^2
+(4c_2^{[2]}+9Dc_5^{[2]})L_{-1}W_{-1}\nn\\
&\quad&+(4c_1^{[2]}+2c_2^{[2]}+\f{9}{2}Dc_5^{[2]})W_{-2}+2c_3^{[2]}W_{-1}^2]|v_0\>.\label{w0onv}\eeq
It turns out that the product $\<v_k|\wh{v}_k\>$ is two times an
element of the inverse Gram/Shapovalov matrix. And $\<v_k|W_0|v_k\>$
is proportional to the $k$-instanton contribution to
$<\m{tr}\varphi^3>$ of gauge theory, up to a factor. The direct
calculation gives \beq
\<v_1|W_0|v_1\>&=&\f{2w(\Delta-3)}{9(D\Delta^2-w^2)}=w\widetilde{Z}_1+2K_{[1]}^{-1}([1],\emptyset;\emptyset,[1])\nn\\
&=&-2(\f{6^3}{22+5c})^{1/2}(-\epsilon_1\epsilon_2)^{-3/2}(\epsilon_1\epsilon_2)^3(\f{22+5c}{6^3})\m{tr}\varphi^3_{k=1},\\
\<v_2|W_0|v_2\>&=&w\widetilde{Z}_2+2K_{[2]}^{-1}([1],[1];\emptyset,[1^2])\nn\\
&=&-2(\f{6^3}{22+5c})^{1/2}(-\epsilon_1\epsilon_2)^{-3/2}(\epsilon_1\epsilon_2)^6(\f{22+5c}{6^3})^2\m{tr}\varphi^3_{k=2},\\
\<v_3|W_0|v_3\>&=&wK_{[3]}^{-1}(\emptyset,[1^3];\emptyset,[1^3])+2K_{[3]}^{-1}([1],[1^2];\emptyset,[1^3]).\eeq

What about higher level? The Gaiotto state $|v_k\>$ can be
constructed by: \be |v_k\>=\sum
c_{Y_1,Y_2}^{[k]}L_{-Y_1}W_{-Y_2}|v_0\>=\sum
K_{[k]}^{-1}(\emptyset,[1^k];Y_1,Y_2)L_{-Y_1}W_{-Y_2}|v_0\>,\ee and
based on the observation of the first three level results, we
conjecture they satisfy\beq
\<v_k|v_k\>&=&K_{[k]}^{-1}(\emptyset,[1^k];\emptyset,[1^k])=(\epsilon_1\epsilon_2)^{3k}(\f{22+5c}{6^3})^kZ_k,\\
\<v_k|W_0|v_k\>&=&wK_{[k]}^{-1}(\emptyset,[1^k];\emptyset,[1^k])+2K_{[k]}^{-1}([1],[1^{k-1}];\emptyset,[1^k])\nn\\
&=&-2(\f{6^3}{22+5c})^{1/2}(-\epsilon_1\epsilon_2)^{-3/2}\left((\epsilon_1\epsilon_2)^{3}(\f{22+5c}{6^3})\right)^k\m{tr}\varphi^3_k.\eeq
Here $K_{[k]}^{-1}(\emptyset,[1^k];\emptyset,[1^k])$ is related to
states $W_{-1}^k|v_0\>$ and its conjugate;
$K_{[k]}^{-1}([1],[1^{k-1}];\emptyset,[1^k])$ is related to
$W_{-1}^k|v_0\>$ and conjugate of $L_{-1}W_{-1}^{k-1}|v_0\>$. Then
we can express the expectation value $<\m{tr}\varphi^3>$ in terms of
CFT data.

In order to rewrite the above relations as (\ref{parti}) and
(\ref{hamil}), we need to scale $|v_k\>$ to absorb the factor
$(\epsilon_1\epsilon_2)^{3k}(\f{22+5c}{6^3})^k$, and scale $W_0$ to
absorb the factor
$-2(\f{6^3}{22+5c})^{1/2}(-\epsilon_1\epsilon_2)^{-3/2}$, then we
have $\<v_k|v_k\>=Z_k$ and $\<v_k|W_0|v_k\>=\m{tr}\varphi^3_k$.

\section{Torus one point correlator}

For the $A_2$ Toda CFT, its one point correlator on the torus
$\<V_m(1)\>_{g=1}$ is given by\be \<V_m(1)\>_{g=1}=\int d\alpha
C(\alpha,\alpha_m,2Q-\alpha)|q^{\Delta_\alpha}\mathcal{F}_{\alpha}^{\
m}(q)|^2,\ee where $C(\alpha,\alpha_m,2Q-\alpha)$ is the structure
constant of three point function, and $\alpha$ is the  momentum of
intermediate channel state, $\alpha_m$ is the momentum of external
state. We use $q$ as the expansion parameter for CFT because, as
will be clear later, it is the same as the instanton expansion
parameter of $\mathcal{N}=2^*$ gauge theory. The one point conformal
block is\be \mathcal{F}_{\alpha}^{\
m}(q)=\sum_{\wt{Y}_{1,2}Y_{1,2}}K^{-1}(\wt{Y}_1,\wt{Y}_2;Y_1,Y_2)\f{\<
L_{-\wt{Y}_1}W_{-\wt{Y}_2}v_0(0)|V_m(1)|L_{-Y_1}W_{-Y_2}v_0(\infty)\>}{\<
v_0(0)|V_m(1)|v_0(\infty)\>}q^{|Y_1|+|Y_2|}.\label{1ptblock}\ee
where $V_m(1)$ is the external state related to mass deformation of
gauge theory. According to the proposal of Wyllard\cite{wyllard}, in
order to establish the AGT relation for the $W_3$ CFT, the momentum
$\alpha$ takes generic value and $\Delta, w$ associated to it are
identified with gauge theory parameters as in(\ref{paraident}), the
external state $V_m(1)$ should be constrained by the semi-null
condition\be (L_{-1}-\f{2\Delta_m}{3w_m}W_{-1})|V_m(1)\>=0.\ee With
this condition, $C(\alpha,\alpha_m,2Q-\alpha)$ is known\cite{fl0505,
fl0709}, and higher point correlators of primary fields can be
evaluated through the three point correlators, therefore solve the
$W_3$ CFT under this condition. Apply $W_1$ on this condition we
have $D_m\Delta_m^2-w_m^2=0$. We find the following appropriate
identification,\be \Delta_m=\f{3m(Q-m)}{\epsilon_1\epsilon_2},\qquad
w_m=[\f{32}{22+5c}(\Delta_m+\f{1}{5})-\f{1}{5}]^{\f{1}{2}}\Delta_m.\ee
We need the vertex $\Gamma(\wt{Y}_{1,2},\emptyset,Y_{1,2})=\<
L_{-\wt{Y}_1}W_{-\wt{Y}_2}v_0(0)|V_m(1)|L_{-Y_1}W_{-Y_2}v_0(\infty)\>$
to compute $\mathcal{F}_{\alpha}^{\ m}(q)$, see\cite{mmmm} for the
CFT technique about it. The first few level vertexes are given
in\cite{kms}. As the elements of the (inverse) Gram/Shapovalov
matrix is only nonzero for $|\wt{Y}_1|+|\wt{Y}_2|=|Y_1|+|Y_2|$, only
the vertex at the level $[1,1]$ in\cite{kms} is useful to check the
torus one point block.

When we compare the conformal block with the instanton partition
function, as the AGT\cite{agt} demonstrated, there is an U(1) factor
presented. For the case of $W_3$ CFT, the relation should be \be
Z^{inst}_{SU(3)}(a,m,q)=[\prod_{i=1}^{\infty}(1-q^i)]^{\f{3m(Q-m)}{\epsilon_1\epsilon_2}-1}
\mathcal{F}_{\alpha}^{\ m}(q).\label{adjointrelat}\ee It is easy to
verify this for $k=1$ with parameter
identification(\ref{paraident}).

Then we may consider how to realize $\<G|W_0^{(s)}|G\>$ in the CFT
conformal block. When we derive the 1-point conformal block
(\ref{1ptblock}) we actually sew two legs of the same pants. Now let
us twist a leg by the operator $W_0^{(s)}$ first, then sew the leg
with an untwisted leg.

Denote the $L_0$ twisted sewing,\be L_\alpha^{\
m}(q)=\sum_{\wt{Y}_{1,2}Y_{1,2}}K^{-1}(\wt{Y}_1,\wt{Y}_2;Y_1,Y_2)\f{\<
L_{-\wt{Y}_1}W_{-\wt{Y}_2}v_0(0)|V_m(1)|L_0L_{-Y_1}W_{-Y_2}v_0(\infty)\>}{\<
v_0(0)|V_m(1)|v_0(\infty)\>}q^{|Y_1|+|Y_2|},\ee it is easy to see\be
\f{L_\alpha^{\ m}(q)}{\mathcal {F}_\alpha^{\
m}(q)}=\Delta+q\f{\p}{\p q}\ln \mathcal {F}_\alpha^{\ m}(q).\ee Then
we can relate $<\m{tr}\varphi^2>$ of $\mathcal{N}=2^*$ theory and
CFT data as \beq (-\epsilon_1\epsilon_2)^{-1}<\m{tr}\varphi^2>
&=&(\Delta-\f{Q^2}{\epsilon_1\epsilon_2})+q\f{\p}{\p q}\ln
Z^{inst}\nn\\&=&(1-\f{3m(Q-m)}{\epsilon_1\epsilon_2})\f{1-E_2(q)}{24}-\f{Q^2}{\epsilon_1\epsilon_2}
+\f{L_\alpha^{\ m}(q)}{\mathcal {F}_\alpha^{\
m}(q)}.\label{adjtrphi2cft}\eeq We have used the fact about the
Einstein series $E_2(q)=1-24\sum_{i=1}^{\infty}iq^i/(1-q^i)$.

Then consider the $W_0$ twisted sewing,\be W_\alpha^{\
m}(q)=\sum_{\wt{Y}_{1,2}Y_{1,2}}K^{-1}(\wt{Y}_1,\wt{Y}_2;Y_1,Y_2)\f{\<
L_{-\wt{Y}_1}W_{-\wt{Y}_2}v_0(0)|V_m(1)|W_0L_{-Y_1}W_{-Y_2}v_0(\infty)\>}{\<
v_0(0)|V_m(1)|v_0(\infty)\>}q^{|Y_1|+|Y_2|},\ee $W_\alpha^{\ m}(q)$
would be more complex than $L_\alpha^{\ m}(q)$, we do not obtain a
general result for arbitrary level $k$, but we observe the following
relation for the first level, \be
-2(\f{6^3}{22+5c})^{1/2}(-\epsilon_1\epsilon_2)^{-3/2}\left(\m{tr}\varphi^3_{k=1}-\f{3}{2}Q(m-\epsilon_1)(m-\epsilon_2)\right)
=[\prod_{i=1}^{\infty}(1-q^i)]^{\f{3m(Q-m)}{\epsilon_1\epsilon_2}-1}W_\alpha^{\
m}(q)|_{k=1},\ee and obviously the following relation also holds,\be
-2(\f{6^3}{22+5c})^{1/2}(-\epsilon_1\epsilon_2)^{-3/2}\m{tr}\varphi^3_{k=0}
=[\prod_{i=1}^{\infty}(1-q^i)]^{\f{3m(Q-m)}{\epsilon_1\epsilon_2}-1}W_\alpha^{\
m}(q)|_{k=0}.\ee Without several higher level data we are unable to
determine the general relation, but we expect on the left hand side
there are new terms come from an U(1) factor similar to the terms on
the right hand side in (\ref{adjtrphi2cft}). The important point is
that this factor involves $Q(m-\epsilon_1)(m-\epsilon_2)$ and is
independent of $a_i$.

\section{Conclusion}

Since the AGT proposal\cite{agt}, there has appeared some strong
evidences that the $\mathcal{N}=2$ gauge theory in the $\Omega$
background has the $W$ symmetry. In this paper we provide an
observation, through the SU(3) gauge theory and $W_3$ algebra, that
the Casimirs of gauge theory can be obtained from the CFT data,
consistent with the form of Seiberg-Witten curve. Hopefully, this
would be true for general cases. The $W_N$ algebra contains the
commutation $[L_m, W^{(s)}_n]=((s-1)m-n)W^{(s)}_{m+n}$ for $s\ge3$,
so $[L_0, W^{(s)}_0]=0$, this fact supports the expectation that
$W^{(s)}_0$ would give all other Hamiltonian as in(\ref{hamil}), up
to factors independent of the quasimomenta. However, it is very hard
to demonstrate the details for general case along this way because
the full commutation relations for $W_N$ algebra would be very
complicated. Although there exists the free field realization of the
$W_N$ algebra, the higher spin currents can be constructed from free
fields through the quantum Miura transform, and in fact some
properties relevant for AGT can be obtained from this
construction\cite{kmst, sv}, however, a general treatment that
incorporate the $W_N$ symmetry into the gauge theory context has not
been presented. Therefore, an understanding of the $W$ symmetry in
$\mathcal{N}=2$ theory from a more fundamental level is surely
desired. On the physics side, it is the mysterious six dimensional
(0, 2) superconformal theory that inspired some recent progress on
$\mathcal{N}=2$ gauge theory\cite{g0904, agt}, hence it might be
worth to pursue an explanation from six dimensional perspective, see
some attempts in\cite{1112, 1205}.

We can also consider how the integrable hierarchy of $\mathcal{N}=2$
gauge theory\cite{mn} can be realized in CFT. Turning on higher
Casimirs in the gauge theory Lagrangian results in the deformed
partition function $Z(\vec{t},a, q)$ which can be evaluated by the
localization method. The final partition function is just
multiplying the N-tuple Young diagram contribution of the undeformed
theory by a factor $\exp(\sum t_m\m{ch}_m(\lbrace
Y_\alpha\rbrace))$\cite{mn, ny0311}. Naively, we may think on the
CFT side this is achieved by considering the correlator
$\<G|\exp(\sum t_sW_0^{(s)})|G\>$. But this does not work. The
operator $L_0$ is diagonal in each subspace of level $k$ in the
Verma module, therefore can be ``exponentialized", but from
(\ref{w0onv}) we know that $W_0$ and zero modes of higher spin
currents are not diagonal in the subspace. The first few level
calculation for the $W_3$ CFT shows $\<G|\exp(t_3W_0)|G\>$ and
$Z(t_3, a, q)$ are not equal(turning on $t_2$ only trivially shift
the gauge coupling $\tau$). It would be interesting to make this
clear.

\section*{Acknowledgments}

I thank Institute of Modern Physics at Xibei University, especially
Prof. Wen-Li Yang, for hospitality during a visit. I also thank
Prof. Hong L\"{u} for correspondence. This work is partially
supported by NSFC No. 11031005.

\section{Appendix: Gram/Shapovalov matrix}\label{appendixI}
The Gram matrix of Shapovalov form is a symmetric sesquilinear form
defined on the Verma module, it is block diagonal with each block
corresponds to the level $k$ subspace of the Verma module. So we
discuss the matrix at each level separately. At the level $k$ it is
denoted by $K_{[k]}$, its elements are denoted by $K_{[k]ij}$, we
denote the cofactor matrix of $K_{[k]}$ by $\widetilde{K}_{[k]}$.
Then the inverse Gram/Shapovalov matrix is denoted by
$K_{[k]}^{-1}$, and its elements are
$K_{[k]ij}^{-1}=\f{\widetilde{K}_{[k]ij}}{\m{det}(K_{[k]})}$.

For SU(2) gauge theory, the corresponding CFT is Liouville CFT with
Virasoro symmetry. At the level $k=1$, the Gram/Shapovalov matrix is
one dimensional, $K^{A_1}_{[1]}=2\Delta$. At the level $k=2$, it is
\begin{equation} K^{A_1}_{[2]}=\left( \begin{matrix}
4\Delta+\f{c}{2} & 6\Delta \\
6\Delta & 4\Delta(2\Delta+1)
\end{matrix}\right)
\end{equation}
At the level $k=3$, it is \begin{equation} K^{A_1}_{[3]}=\left(
\begin{matrix}
6\Delta+2 & 10\Delta & 24\Delta \\
10\Delta & 8\Delta(\Delta+1)+c\Delta & 12\Delta(3\Delta+1)\\
24\Delta & 12\Delta(3\Delta+1) & 24\Delta(\Delta+1)(2\Delta+1)
\end{matrix}\right)
\end{equation}

For SU(3) gauge theory, the corresponding CFT is $A_2$ Toda CFT with
$W_3$ symmetry. At the level $k=1$, denote the base vector
$L_{-1}\rp, W_{-1}\rp$ by $|i\>, i=1,2$, then we have
\begin{equation}
K^{A_2}_{[1]}=\left( \begin{matrix}
2\Delta & 3w \\
3w & \f{9}{2}D\Delta
\end{matrix}\right)
\end{equation}

At the level $k=2$ denote the base vector $L_{-2}\rp, L_{-1}^2\rp,
L_{-1}W_{-1}\rp, W_{-2}\rp, W_{-1}^2\rp$ by $|i\>, i=1,2,3,4,5$,
then we have
\begin{equation}
K^{A_2}_{[2]}={\small \left( \begin{matrix}
4\Delta+\f{c}{2} & 6\Delta & 9w & 6w & \f{45}{2}D\Delta \\
6\Delta & 4\Delta(2\Delta+1) & 6w(2\Delta+1) & 12w & 27D\Delta+18w^2 \\
9w & 6w(2\Delta+1) & 9D\Delta^2+9D\Delta+9w^2 & 18D\Delta & \f{27}{2}Dw(2\Delta+3) \\
6w & 12w & 18D\Delta & 9\Delta(D+1) & \f{27}{2}w(3D+1) \\
\f{45}{2}D\Delta & 27D\Delta+18w^2 & \f{27}{2}Dw(2\Delta+3) &
\f{27}{2}w(3D+1) &
\f{81}{4}D^2\Delta(2\Delta+1)+\f{648D\Delta(\Delta+1)+4w^2}{22+5c}
\end{matrix}\right)}
\end{equation}
In this notation,
$K_{[2]}(\emptyset,[1^2];\emptyset,[1^2])=K_{[2]55}$, and
$K_{[2]}([1],[1];\emptyset,[1^2])=K_{[2]35}$. As has been shown
in\cite{mm} that $\m{det}(K_{A_2}^{[2]})$ is factorizable, related
to the gauge theory expression through the
identification(\ref{paraident}), \be
\m{det}(K^{A_2}_{[2]})=2^{16}3^8(22+5c)^{-4}(\epsilon_1\epsilon_2)^{-12}\prod_{\alpha<\beta}^3(a_{\alpha\beta}^2-\epsilon_+^2)^2(a_{\alpha\beta}^2-(\epsilon_++\epsilon_1)^2)(a_{\alpha\beta}^2-(\epsilon_++\epsilon_2)^2).\ee
A useful fact for study $\<v_2|W_0|v_2\>$ is the element
$K_{[2]35}^{-1}$, as we have \beq
\wt{K}_{[2]35}^{A_2}&=&2^43^5(22+5c)^{-2}w(\Delta-1)(22w^2+5cw^2-2\Delta^2+c\Delta^2-32\Delta^3)\nn\\
&\quad&(2c+c^2-44w^2-10cw^2-28\Delta-12c\Delta+c^2\Delta+40\Delta^2+16c\Delta^2+64\Delta^3),\eeq
therefore write in the gauge theory parameters we have\be
K_{[2]35}^{-1}=-(\epsilon_1\epsilon_2)^2(\f{-4\epsilon_1\epsilon_2-15Q^2}{3})^{3/2}\f{a_1a_2(a_1+a_2)(a_1^2+a_1a_2+a_2^2-\epsilon_+^2+\epsilon_1\epsilon_2)
(8a_1^6+24a_1^5a_2+\cdots)}{\prod_{\alpha<\beta}(a_{\alpha\beta}^2-\epsilon_+^2)(a_{\alpha\beta}^2-(\epsilon_++\epsilon_1)^2)(a_{\alpha\beta}^2-(\epsilon_++\epsilon_2)^2)}\ee
where the abbreviated polynomial is of degree six and is the same as
that appears in $\m{tr}\varphi^3_{k=2}$ for pure gauge theory:
$8a_1^6+24a_1^5a_2-6a_1^4a_2^2+\cdots-2160\epsilon_1^2\epsilon_2^4-825\epsilon_1\epsilon_2^5-128\epsilon_2^6.$

\end{document}